\RequirePackage{amsmath}
\documentclass[runningheads]{llncs}
\usepackage{graphicx}
\begin{document}
	\title{Adaptive Chemotaxis for improved Contour Tracking using Spiking Neural Networks}
	%
	%
	\author{Shashwat Shukla \inst{1}\textsuperscript{$\dagger$} \and
		Rohan Pathak \inst{1}\textsuperscript{$\dagger$} \and
		Vivek Saraswat \inst{1} \and 
		Udayan Ganguly \inst{1}}
	%
	%
	\institute{Department of Electrical Engineering, IIT Bombay\\
		\email{shashwat.shukla@iitb.ac.in} \\
		$\dagger$: Equal contribution}
	\maketitle              
	\begin{abstract}
		In this paper we present a Spiking Neural Network (SNN) for autonomous navigation, inspired by the chemotaxis network of the worm Caenorhabditis elegans. In particular, we focus on the problem of contour tracking, wherein the bot must reach and subsequently follow a desired concentration setpoint. Past schemes that used only klinokinesis can follow the contour efficiently but take excessive time to reach the setpoint. We address this shortcoming by proposing a novel adaptive klinotaxis mechanism that builds upon a previously proposed gradient climbing circuit. We demonstrate how our klinotaxis circuit can autonomously be configured to perform gradient ascent, gradient descent and subsequently be disabled to seamlessly integrate with the aforementioned klinokinesis circuit. We also incorporate speed regulation (orthokinesis) to further improve contour tracking performance. Thus for the first time, we present a model that successfully integrates klinokinesis, klinotaxis and orthokinesis. We demonstrate via contour tracking simulations that our proposed scheme achieves an 2.4x reduction in the time to reach the setpoint, along with a simultaneous 8.7x reduction in average deviation from the setpoint.
	\end{abstract}
	\section{Introduction}
	
	The worm Caenorhabditis elegans (C. elegans) is a model organism for neurobiology as it displays fairly sophisticated behavior despite having only 302 neurons. One such behavior of interest is chemotaxis: the ability to sense chemicals such as NaCl and to then move in response to the sensed concentration. 
	The worm prefers certain concentrations of NaCl as it associates them with finding food, with these concentrations thus acting as setpoints for the worm. 
	This ability to search for and follow the level set (which is an isocontour in 2D) for a particular setpoint concentration is called contour tracking, and has been observed experimentally in the worm \cite{luo2006sensorimotor}. Remarkably, the worm is able to track contours in a highly resource-constrained manner with just one concentration sensor and a small number of neurons. Contour tracking is also an important function for autonomously navigating robots, and it is thus of interest from an engineering standpoint to study the small yet efficient chemotaxis circuit of C. elegans. The emergence of energy-efficient nanoscale Neuromorphic hardware \cite{dutta2017leaky} motivates mapping these compact chemotaxis circuits onto Spiking Neural Networks (SNNs) in order to instantiate autonomously navigating robots operating under severe resource and energy constraints.
	
	One of the strategies that the worm uses is called klinokinesis, wherein the worm makes abrupt turns away from its current direction. Klinokinesis requires the worm to compare the current sensed concentration with past samples to estimate the concentration gradient along its path of motion. It turns away when it is above the setpoint and senses a positive gradient, or if it is below the setpoint and senses a negative gradient. It thus corrects its path so that it is always moving towards the setpoint. 
	A model for the sensory neurons used to compute temporal derivatives was proposed in \cite{appleby2012model}. 
	Santurkar and Rajendran added motor neurons to the model from \cite{appleby2012model} to propose an SNN for klinokinesis \cite{santurkar2015c}. They also demonstrated hardware compatibility with standard CMOS circuitry. However, their SNN required external currents to operate correctly and thus was not an autonomous solution. Shukla, Dutta and Ganguly resolved this problem by designing SNNs for the rate-coded logic operations required by the klinokinesis circuit and ensuring correctness of operation \cite{shukla2018design}. Furthermore, they incorporated an additional sub-circuit to allow the worm to escape local extrema, reduced the response latency of the SNN by incorporating anticipatory control and demonstrated feasibility on nanoscale neuromorphic hardware. 
	
	An important limitation of klinokinesis is that it only uses the sign of the gradient and not its magnitude. Thus while klinokinesis ensures that the worm is always moving towards the setpoint, it does not ensure that it takes the shortest path. By definition, the direction with the highest gradient magnitude corresponds to the path of steepest ascent or descent. Indeed, the worm is known to align itself along (or against) the gradient via gradual turns in a process called klinotaxis. The worm performs klinotaxis by estimating the spatial gradient in the direction perpendicular to its current path by comparing concentration values to the left and right of its head while moving in a snake-like sinusoidal motion, using this estimate to gradually correct its path. Izquierdo and Lockery proposed a mechanistic model for gradient ascent using klinotaxis and learned model parameters via evolutionary algorithms \cite{izquierdo2010evolution}. Izquierdo and Beer subsequently attempted to map this model onto the connectome of the worm \cite{izquierdo2013connecting}. 
	
	\begin{figure}[htbp]
		\centerline{\includegraphics[width=0.85\linewidth]{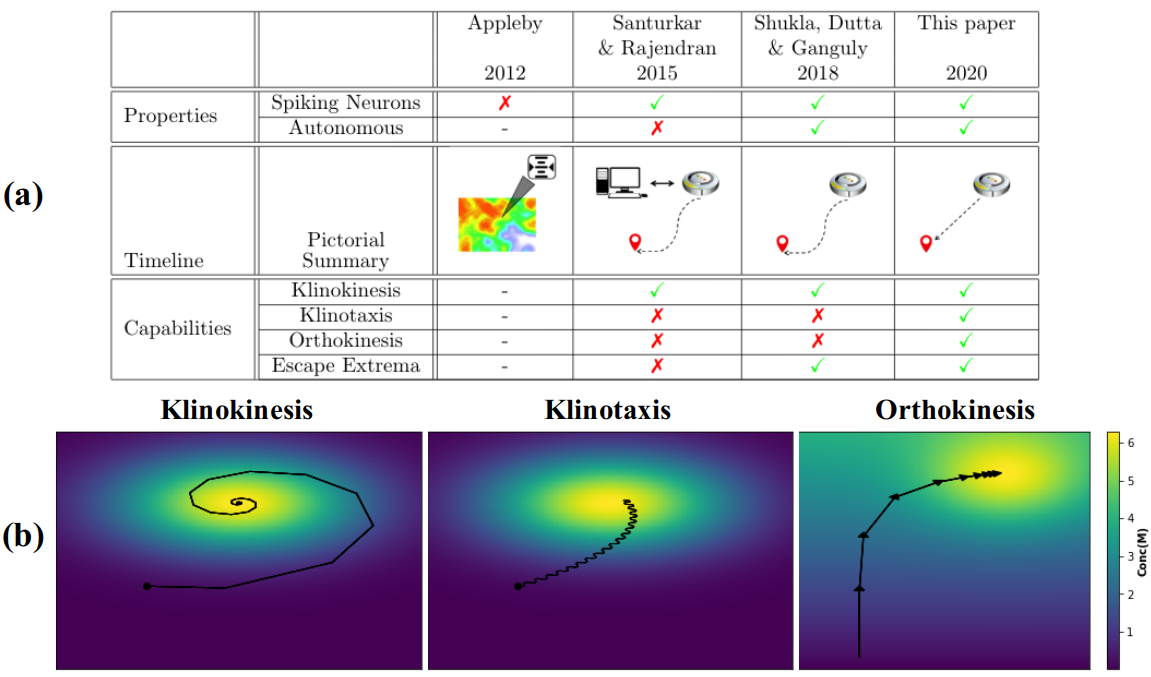}}
		\caption{(a) Timeline and comparison of this paper with past literature. (b) Gradient ascent to illustrate navigation mechanisms. Left: The bot makes abrupt turns to correct its path by using only the sign of the gradient, and thus takes a circuitous route to the peak. Center: The bot gradually corrects its path to align with the direction of steepest ascent and thus takes a much shorter route to the peak. Also note the sinusoidal motion of the bot. Right: This plot is only to visualize orthokinesis. Note that the arrows become denser close to the peak, depicting how the bot slows down near the setpoint and regions with large gradients.}
		\label{toon}
	\end{figure}
	
	The first important contribution of this paper is to build upon the gradient ascent circuit in \cite{izquierdo2010evolution} to develop a novel adaptive klinotaxis circuit that can be autonomously configured to perform gradient ascent, gradient descent, and disabled upon reaching the setpoint. Second, we implement this adaptive klinotaxis circuit with spiking neurons and then integrate it with the klinokinesis SNN from \cite{shukla2018design}. It is important to note that these strategies serve complementary roles, with klinokinesis allowing for rapid turns to ensure that the worm always moves closer to the setpoint, and klinotaxis allowing the worm to gradually optimize its path \cite{iino2009parallel,itskovits2018concerted}. Indeed, the worm is known to use klinokinesis and klinotaxis in tandem \cite{iino2009parallel}. 
	However, in the context of contour tracking, it is important to understand how klinotaxis and klinokinesis can work together. In particular, the worm must align with the gradient until it reaches the setpoint and must subsequently move perpendicular to the gradient to follow the setpoint contour, thus requiring the worm to change its behavior based on how close it is to the setpoint. This problem was previously encountered in work by Skandari, Iino and Manton who proposed a non-adaptive, non-spiking model that attempts to extend Lockery's work to perform contour tracking \cite{skandari2016analogue}. Their simulations show that their network for reaching the setpoint and their network for subsequently following the contour are incompatible with each other, leading to a failure in tracking contours near regions with large gradients. The adaptive nature of our klinotaxis circuit allows us to address this important problem. Furthermore, the gradual nature of klinotaxis steering leads to large deviations from the setpoint while following the desired contour, a problem that we are able to address by also including the klinokinesis circuit to enable faster turns. Our circuit thus allows us to seamlessly integrate the benefits of both these navigation strategies.
	
	We also incorporate orthokinesis in our SNN model, wherein the bot can also regulate its speed as a function of sensed concentration \cite{benhamou1989animals}. This allows it to slow down near the setpoint and near regions with large gradients, leading to a further reduction in deviations from the setpoint while following the desired contour. To the best of our knowledge, this is the first circuit model that successfully integrates klinokinesis, klinotaxis and orthokinesis.

	\section{Proposed Algorithm}
	
	\subsection{Adaptive Klinotaxis}
	
	Klinotaxis is the mechanism whereby, as the worm moves along its sinusoidal trajectory, it compares the values of sensed concentrations in one half-cycle to those in the next half-cycle, and then changes its course based on this information. Klinotaxis has typically been studied in the context of gradient ascent, wherein the worm will bias its motion towards the side which is better aligned with the local gradient direction, thus gradually aligning with the gradient and performing steepest ascent, and thus allowing the worm to reach the peak faster. Crucially, if the worm had two spatially separated concentration sensors, then it could compare the values from these sensors to estimate the gradient direction, which is called tropotaxis. However, the worm only has one concentration sensor, thus requiring it to use its own body motion to sample to the left and right of its path, as it does with its sinusoidal motion. Such a setting is of great interest for highly resource constrained bots that are too small to carry two bulky sensors and where the spatial separation between sensors is too small to enable tropotaxis.
	In this paper, we enable our bot to not only perform gradient ascent, but also enable gradient descent and the ability to switch off the klinotaxis mechanism entirely. Furthermore, this change in behavior is affected autonomously based on sensed concentration, and we thus call this adaptive klinotaxis. 
	
	\begin{figure}[htbp]
		\centerline{\includegraphics[width=0.7\linewidth]{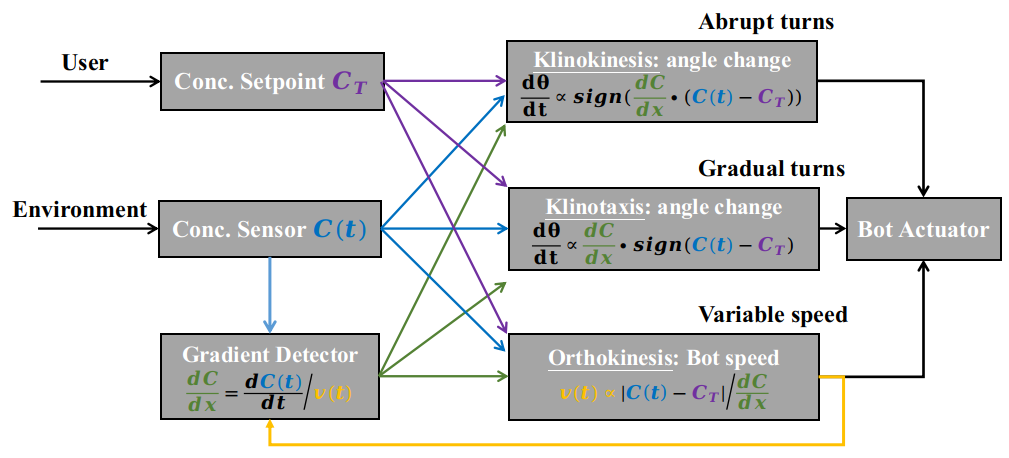}}
		\caption{Block Diagram for the full network.}
		\label{net}
	\end{figure}

	\begin{figure}[htbp]
		\centerline{\includegraphics[width=\linewidth]{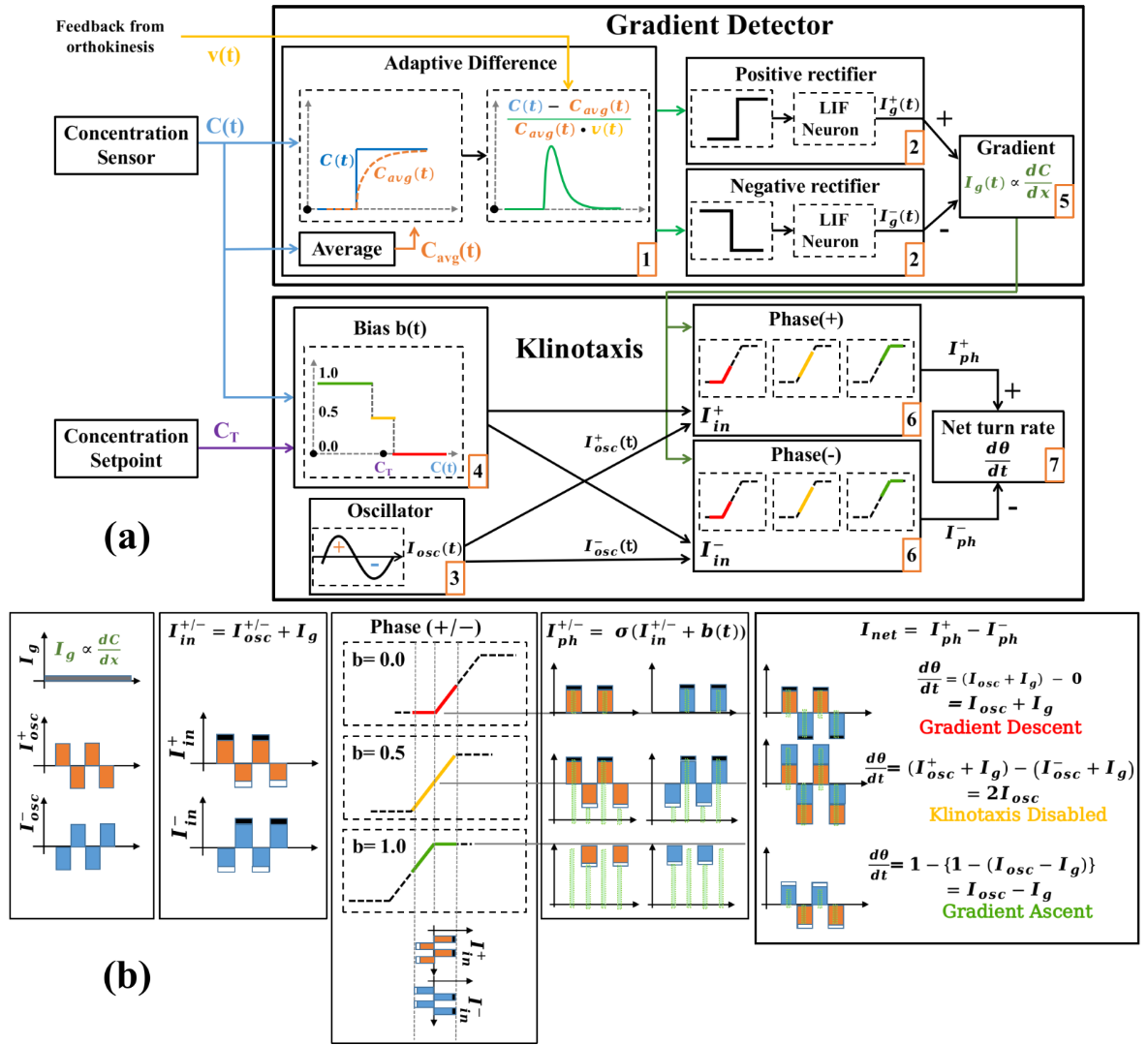}}
		\caption{(a) Network diagram for the gradient detector and klinotaxis blocks. The gradient detector functions by computing a temporal estimate for the gradient: $C(t) - C_{avg}(t)$, which is normalized by $C_{avg}(t)$ to make this estimate scale-invariant. It is further divided by $v(t)$ to convert the temporal derivative to a spatial derivative ($\frac{\mathrm{d} C}{\mathrm{d} t} = v(t) \cdot \frac{\mathrm{d} C}{\mathrm{d}x}$). This spatial derivative is then rate-coded using two sparsely firing leaky-integrate-and-fire neurons, with the refractory periods acting as saturating non-linearity. (b) Signal flow through the klinotaxis block. The bias shifts the piecewise linear sigmoidal response curve to the left or right, while the sum of currents from the oscillator and gradient detector, $I^{\pm}_{ph}(t)$, is the input to these shifted response curves. For $b=0.0$: The positive parts of the input are retained and thus the turning rate increases with $I_g(t)$. Thus the bot turns slower when aligned against the gradient - gradient descent. For $b=0.5$: Both the positive and negative parts of the input are retained and the effect of $I_g(t)$ is cancelled out. Thus the bot's turning rate is independent of the gradient - klinotaxis disabled. For $b=1.0$: Effectively, the negative parts of the input are retained and thus the turning rate decreases with $I_g(t)$. Thus the bot turns slower when aligned along the gradient - gradient ascent.}
		\label{taxnet}
	\end{figure}
	
	The gradient detector and klinotaxis blocks are depicted in Fig. \ref{taxnet}(a). Like the worm, our bot has a single concentration sensor whose output at time $t$ is the sensed concentration $C(t)$. This $C(t)$ is used to compute an adaptive difference-estimate for the gradient, $I_{ad}(t)$, given as:
	
	\begin{equation}
	I_{ad}(t) = \frac{C(t) - C_{avg}(t)}{C_{avg}(t) \cdot v(t)} \\
	\end{equation}
	
	Here $C_{avg}(t)$ is the sensed concentration averaged over the past $10$ seconds, and thus $C(t) - C_{avg}(t)$ is an estimate for the temporal derivative $\frac{\mathrm{d} C}{\mathrm{d} t}$. This is dynamically scaled by $\frac{1}{C_{avg}(t)}$ to allow the gradient detector neurons to effectively utilize available signaling bandwidth and to make its response invariant to the average concentration, allowing our bot to operate in environments where the concentrations can vary over many orders of magnitude. This is thus a simple but important modification to the static scaling used in \cite{santurkar2015c,shukla2018design,izquierdo2010evolution,izquierdo2013connecting}. This is in turn scaled by $\frac{1}{v(t)}$ to convert the temporal derivative to a spatial derivative, in line with $\frac{\mathrm{d} C}{\mathrm{d} x} = \frac{1}{v(t)} \cdot \frac{\mathrm{d} C}{\mathrm{d} t}$. This requires the existence of a feedback loop which gives the sensory neurons access to the bot's velocity, as depicted in Fig. \ref{net}.
	Note that such divisive gain modulation has also been observed in neurons in-vivo \cite{bastian1986gain,vestergaard2015divisive}.
	Also note that while operating with a constant speed, as was the case in \cite{santurkar2015c,shukla2018design}, the temporal and spatial gradients are linearly proportional to each other and thus there was no need for such speed-dependent scaling in past work. 	
	
	The positive and negative parts of this signal given are respectively denoted by $I^+_{ad}(t) = I_{ad}(t) \cdot \delta(I_{ad}(t) > 0) ~;~ I^-_{ad}(t) = I_{ad}(t) \cdot \delta(I_{ad}(t) < 0)$, where $\delta(.)$ is a delta-function which is $1$ if the input condition is true and is $0$ otherwise. Next, $I^+_{ad}(t)$ and $I^-_{ad}(t)$ are respectively fed into neurons $N_+$ and $N_-$, noting that $I_{ad}(t)$ was encoded this way using two neurons because neurons can only have positive firing rates. We model $N_+$ and $N_-$ as leaky integrate-and-fire (LIF) neurons with respective membrane potentials $V_+$ and $V_-$ which evolve as:
	
	\begin{equation}
	C_G\frac{\mathrm{d} V_+}{\mathrm{d} t} = I^+_{ad}(t) - \frac{V_+(t)}{R_G} ~;~
	C_G\frac{\mathrm{d} V_-}{\mathrm{d} t} = I^-_{ad}(t) - \frac{V_-(t)}{R_G}
	\end{equation}
	
	Here $C_G$ and $R_G$ are respectively the membrane capacitance and resistance. The neurons $N_+$ and $N_-$ fire when $V_+$ and $V_-$ respectively cross the firing threshold $V_T$, and the membrane voltage is then reset to $0$ for the duration of the refractory period. 
	The spike-trains of $N_+$ and $N_-$ are convolved with the kernel $\kappa(t) = e^{\frac{-t}{\tau_1}} - e^{\frac{-t}{\tau_2}} ~;~ \tau_1 > \tau_2$ to generate the respective output currents $I_g^+(t)$ and $I_g^-(t)$.
	This is an instance of rate-coding wherein $I_g^{\pm}(t)$ increases with $I^{\pm}_{ad}(t)$. However this mapping is non-linear due to the refractory period, and crucially, $I_g^{\pm}(t)$ saturates for large values of $I^{\pm}_{ad}(t)$. 
	The refractory period and parameters of $\kappa(t)$ are chosen so that this maximum value of $I_g^{\pm}(t)$ is $1$, a fact that will be used in the klinotaxis circuit. 
	Apart from this non-linear response, the other advantage of using spiking neurons is that unlike analog neurons they are not always on and are thus much more energy-efficient. 
	Finallly, the rate-coded gradient estimate is given as $I_g(t) = I_g^+(t) - I_g^-(t)$.
	
	Having discussed the gradient detector circuit, we now proceed to describe the klinotaxis circuit. 
	The first component is the oscillator current $I_{osc}(t)$ with time period $T_{osc}$, which is used to generate two oscillatory signals with opposite phases as: $I^+_{osc}(t) = I_{osc}(t)$ and $I^-_{osc}(t) = -I_{osc}(t)$. The second component is the bias function which determines the mode of operation of the klinotaxis circuit and is denoted by $b(t)$. The third input, $I_g(t)$, comes from the gradient detector discussed above.
	Output from these three blocks is fed to the two non-linear ``phase'' blocks, denoted by ``Phase($\pm$)'' in Fig. \ref{taxnet}. These phase blocks are the most important part of the circuit, yielding output currents $I^{\pm}_{ph}(t)$. The net turning rate $\frac{\mathrm{d} \theta(t)}{\mathrm{d} t}$ due to klinotaxis is given by the scaled difference in output of these two phase blocks, where $\theta(t)$ is the bot's steering angle. We choose the convention wherein a positive change in $\theta$ will correspond to turning clockwise.
	
	\begin{align}
	I_{osc}(t) &= I^+_{osc}(t) = - I^-_{osc}(t) = \sin(\frac{2\pi t}{T_{osc}}) \\
	b(t) &= 
	\begin{aligned}
	\left\{\begin{matrix}
	1.0 ~;~ &(C_T - C(t)) > \varepsilon \\ 
	0.5 ~;~ &|C(t) - C_T| < \varepsilon \\ 
	0.0 ~;~ &(C(t) - C_T) > \varepsilon
	\end{matrix}\right.
	\end{aligned}\\
	I_g(t) &= I_g^+(t) - I_g^-(t) \\
	I^{\pm}_{ph}(t) &= \sigma(\alpha \cdot I^{\pm}_{osc}(t) + \beta \cdot I_g(t) + b(t))  \\
	\frac{\mathrm{d} \theta(t)}{\mathrm{d} t} &= w_m \cdot (I^{+}_{ph}(t) - I^{-}_{ph}(t))
	\end{align}
	
	To understand (4), observe that if we wish to reach the setpoint concentration $C_T$ to within an tolerance $\varepsilon$, it is straightforward to see that we want $b(t) = 1.0$ (gradient ascent) for $C(t) < (C_T - \varepsilon)$, $b(t) = 0.0$ for $C(t) > (C_T + \varepsilon)$ (gradient descent), and $b(t) = 0.5$ (disable klinotaxis) for $\left | C(t) - C_T \right | < \varepsilon$. By disabling klinotaxis close to the setpoint, we allow klinokinesis to seamlessly take over, allowing the bot to follow the setpoint contour using klinokinesis as demonstrated in \cite{shukla2018design}. These mechanisms thus serve complementary roles, with klinotaxis used to reach the setpoint, and klinokinesis to subsequently follow the setpoint contour. In (6), the non-linear response of the phase blocks, $\sigma(x)$ (for input $x$), is equal to $0$ for $x < 0$, $x$ for $0<x<1$, and $1$ for $x>1$. Thus it increases linearly between $0$ and $1$ and saturates outside this range. While we have used this piecewise-linear form for $\sigma(x)$ in subsequent analysis, we have used a smoother and more biologically feasible approximation in our final contour tracking simulations, given as $\sigma(x) \approx 0.5 \cdot (1 + \tanh(2 \cdot x - 1))$.  
	Also note that $\alpha$, $\beta$ and $w_M$ are positive scaling constants.
	
\begin{figure}[htbp]
		\centerline{\includegraphics[width=\linewidth]{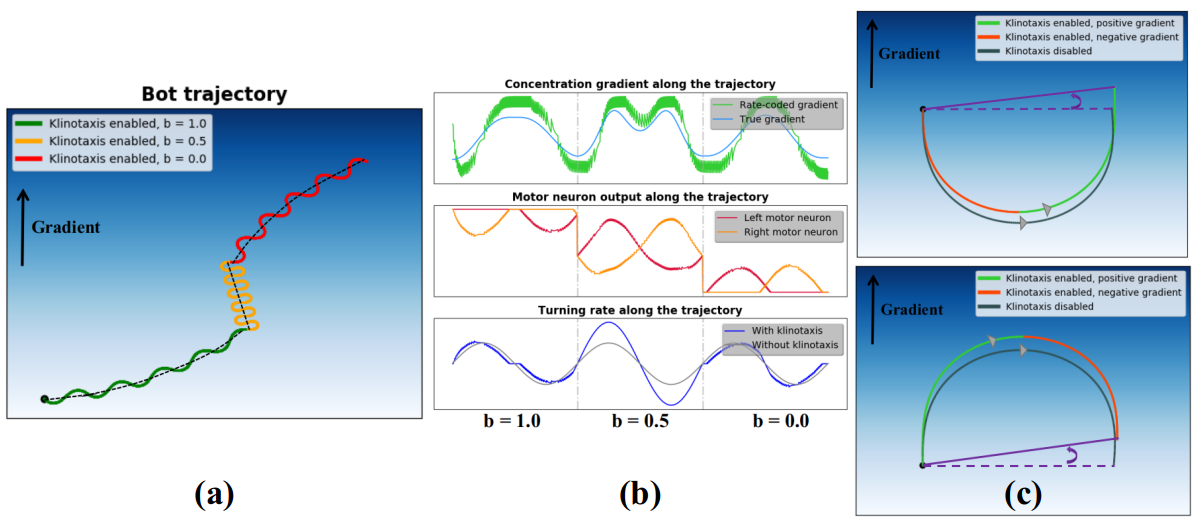}}
		\caption{(a) Path of the bot over five oscillator cycles for the three choices of $b$. Note how the trajectory bends towards, does not bend, and bends away from the steepest gradient direction respectively for $b = \{1.0, 0.5, 0.0\}$. (b) Timeplots for system variables over a represenative cycle, one for each choice of $b$. The top plot shows that the rate-coding estimate is indeed able to follow the true gradient. The middle plot shows the difference in clipping of left ($I^{+}_{ph}(t)$) and right ($I^{-}_{ph}(t)$) motor neurons, for choices of $b$. The bottom plot shows how the gradient changes the turning rate vis-a-vis the zero-gradient case. Note that there is no skew for $b=0.5$, while the skews for $b=0.0$ and $b=1.0$ are precisely opposite to one another. 
			(c) Trajectory of the bot performing gradient ascent for a half-cycle, with and without klinotaxis enabled. Top: The bot is initially aligned against the gradient and is turning clockwise. Bottom: The bot is initially aligned along the gradient and is turning anticlockwise. The green and red parts of the curves respectively correspond to sensing positive and negative gradients, making the bot turn slower and faster respectively. In both cases the axis of motion is initially perpendicular to the gradient and at the end of the half-cycle this axis has rotated clockwise towards the gradient direction when klinotaxis is enabled.}
		\label{taxmech}
	\end{figure}

	We now proceed to describe the adaptive klinotaxis mechanism for the half-cycle from $0$ to $T_{osc} / 2$. We will first do this for $b = 1.0$, corresponding to gradient ascent. In this half-cycle $I^+_{osc}(t)$ is positive and thus $\sigma(1 + \alpha \cdot I^+_{osc}(t)) = 1$, meaning that $I^{+}_{ph}(t)$ saturates to $1$. On the other hand, as $I^-_{osc}(t)$ is negative during this cycle, $\sigma(1 + \alpha \cdot I^-_{osc}(t)) = 1 + \alpha \cdot I^-_{osc}(t)$ and thus $I^{-}_{ph}(t)$ lies in the linear region. Note that these are approximate statements that ignore the contribution from $I_g(t)$. While it can be verified that these statements are exact for $\frac{T_{osc}}{2 \pi} sin^{-1}(\frac{\beta}{\alpha}) < t < \frac{T_{osc}}{2}$ (by noting that both $I_g(t)$ and $I_{osc}(t)$ have maximum amplitude of $1$), they will be assumed for the entire half-cycle for ease of analysis. We are thus interpreting $\beta \cdot I_g(t)$ as a small perturbation to the output. We then have $\frac{\mathrm{d} \theta(t)}{\mathrm{d} t} = 1 - (1 + \alpha \cdot I^-_{osc}(t) + \beta \cdot I_g(t)) = \alpha \cdot (-I^-_{osc}(t)) - \beta \cdot I_g(t) = \alpha \cdot I_{osc}(t) - \beta \cdot I_g(t)$.
	Thus we see that for $b=1.0$ the bot turns slower when aligned along the gradient ($I_g(t) > 0$) and turns faster when aligned against the gradient ($I_g(t)< 0$). The bot's net motion is thus biased and it tends to align along the gradient, thus performing gradient ascent. This is depicted in Fig. \ref{taxmech}(c) wherein the turning rate is higher and lower respectively for the red part ($I_g(t)<0$) and green part ($I_g(t)>0$) of the trajectories. 
	
	Next, we consider gradient descent with $b=0.0$. In the first half-cycle $I^-_{osc}(t)$ is negative and thus $\sigma(0 + \alpha \cdot I^-_{osc}(t)) = 0$, and thus $I^{-}_{ph}(t)$ saturates to $0$. On the other hand, as $I^+_{osc}(t)$ is positive during this cycle, $\sigma(0 + \alpha \cdot I^+_{osc}(t)) = \alpha \cdot I^+_{osc}(t)$ and thus $I^{+}_{ph}(t)$ lies in the linear region. Again, treating $\beta \cdot I_g(t)$ as a perturbation that only affects the non-saturated phase, we have $\frac{\mathrm{d} \theta(t)}{\mathrm{d} t} = (\alpha \cdot I^+_{osc}(t) + \beta \cdot I_g(t)) - 0 = \alpha \cdot I_{osc}(t) + \beta \cdot I_g(t)$. Thus we see that for $b=0.0$ the bot turns faster when aligned along the gradient ($I_g(t) > 0$) and turns slower when aligned against the gradient ($I_g(t)< 0$). The bot's net motion is thus biased and it tends to align against the gradient, thus performing gradient descent. 
	
	Finally, we discuss the case of disabling klinotaxis by setting $b=0.5$. Furthermore, we choose hitherto unspecified constants as $\alpha = 0.4$ and $\beta = 0.1$. For all time $t$ we then have that $\left | 0.1 \cdot I_g(t) \pm 0.4 \cdot I_{osc}(t) + 0.5 \right | < 0.1 + 0.4 + 0.5 = 1$, recalling that by design, both $I_g(t)$ and $I_{osc}(t)$ have a maximum magnitude of $1$. Thus for $b=0.5$, neither $I^{+}_{ph}(t)$ nor $I^{-}_{ph}(t)$ saturate and both lie in the linear region. We then have $\frac{\mathrm{d} \theta(t)}{\mathrm{d} t} = (\alpha \cdot I^+_{osc}(t) + \beta \cdot I_g(t) + 0.5) - (\alpha \cdot I^-_{osc}(t) + \beta \cdot I_g(t) + 0.5) = \alpha \cdot (I^+_{osc}(t) - I^-_{osc}(t))  = 2\alpha \cdot I_{osc}(t)$. 	
	Thus in this case, $I_g(t)$ has no effect on the turning rate of the bot and hence the klinotaxis mechanism stands disabled. 
	We note that the amplitudes of $\alpha = 0.4$ and $\beta = 0.1$ respectively for the oscillatory and gradient terms were chosen such that they sum up to $0.5$. This was done to maximize the dynamic range of input in the linear output regime, while also not allowing this input to saturate. Also note that the larger value of $0.4$ was chosen for the oscillatory component to ensure that the bot swerves to the left and right with a large enough amplitude and thus samples it local environment, despite the modulatory effect of the gradient term. At the same time, the amplitude of $0.1$ for the gradient term is large enough to ensure that it does have a sufficiently large modulatory effect to enable klinotaxis. In summary, the turning rate in the first half-cycle is given as:
	
	\begin{equation}
	\frac{\mathrm{d} \theta}{\mathrm{d} t} = \left\{\begin{matrix}
	w_M \cdot \left (\alpha \cdot I_{osc}(t) + \beta \cdot I_{grad}(t) \right ) ~;~ b = 0.0 \\ w_M \cdot 2\alpha \cdot I_{osc}(t) ~;~ b = 0.5 \\ 
	w_M \cdot \left (\alpha \cdot I_{osc}(t) - \beta \cdot I_{grad}(t) \right ) ~;~ b = 1.0
	\end{matrix}\right.
	\end{equation}
	
	It can be verified that in the next half-cycle from $T_{osc} / 2$ to $T_{osc}$, we get the same expression for $\frac{\mathrm{d} \theta}{\mathrm{d} t}$ as given in (8), but now with a negative sign. 
	Thus the bot turns clockwise in one half-cycle and anti-clockwise in the next. Note also that it suffices to describe one full-cycle as the same mechanism recurs over time.

	\subsection{Klinokinesis}
	
	Klinokinesis is a course correction algorithm, wherein the worm turns around when it senses that it is moving away from the desired setpoint concentration. This happens in two cases: when it senses $\frac{\mathrm{d} C(t)}{\mathrm{d} t} > 0, ~ C(t) > C_T$ or if $\frac{\mathrm{d} C(t)}{\mathrm{d} t} < 0, ~ C(t) < C_T$. 
	Note that this requires computing an AND operation over the sensed gradient and concentration values for which we use the SNN developed in \cite{shukla2018design}. While klinokinesis allows for rapid corrections to the bot's path and is thus well suited to closely following the contour once it is reached, it is not capable of finding the shortest path to the setpoint as it only uses the sign of the gradient and does not seek out the direction of \emph{steepest} descent, thus motivating the inclusion of the complementary mechanism of klinotaxis.
	
	\subsection{Orthokinesis}
	
	We incorporate orthokinesis to reduce overshoot whilst following the setpoint contour using klinokinesis. We describe the bot's speed in discrete-time for ease of understanding, while noting that it is straightforward to convert this to continuous-time. For discrete time $1, ..., t-1, t, ...$, the bot speed is given as:
	
	\begin{equation}
	v[t] = v_c + \frac{k \cdot v[t-1] \cdot \left | C[t] - C_T \right |}{a + \left | \frac{\mathrm{d} C[t-1]}{\mathrm{d} x} \right |}
	\end{equation}
	
	Here $v_c$ is a constant that ensures that the worm continues to move along the setpoint contour despite the second term going to $0$ close to $C_T$. Furthermore, the second term is proportional to $v[t-1]$ as a means of enforcing continuity in the values of $v[t]$. The term $\left | C[t] - C_T \right |$ is included so that the worm slows down close to the setpoint. Thus by allowing the worm to slow down near the setpoint concentration, we enable improved contour tracking. We would also like the bot to slow down near regions with high gradient magnitudes so that it does not overshoot the setpoint. This is ensured by including $\frac{\mathrm{d} C[t-1]}{\mathrm{d} x}$ in the denominator, which is the output of the gradient detector in the previous time step. Finally, $k$ is a constant scaling factor while $a$ is a constant that ensures that the denominator is never $0$.

	\section{Results and Conclusions}
	
\begin{figure}[htbp]
		\centerline{\includegraphics[width=\linewidth]{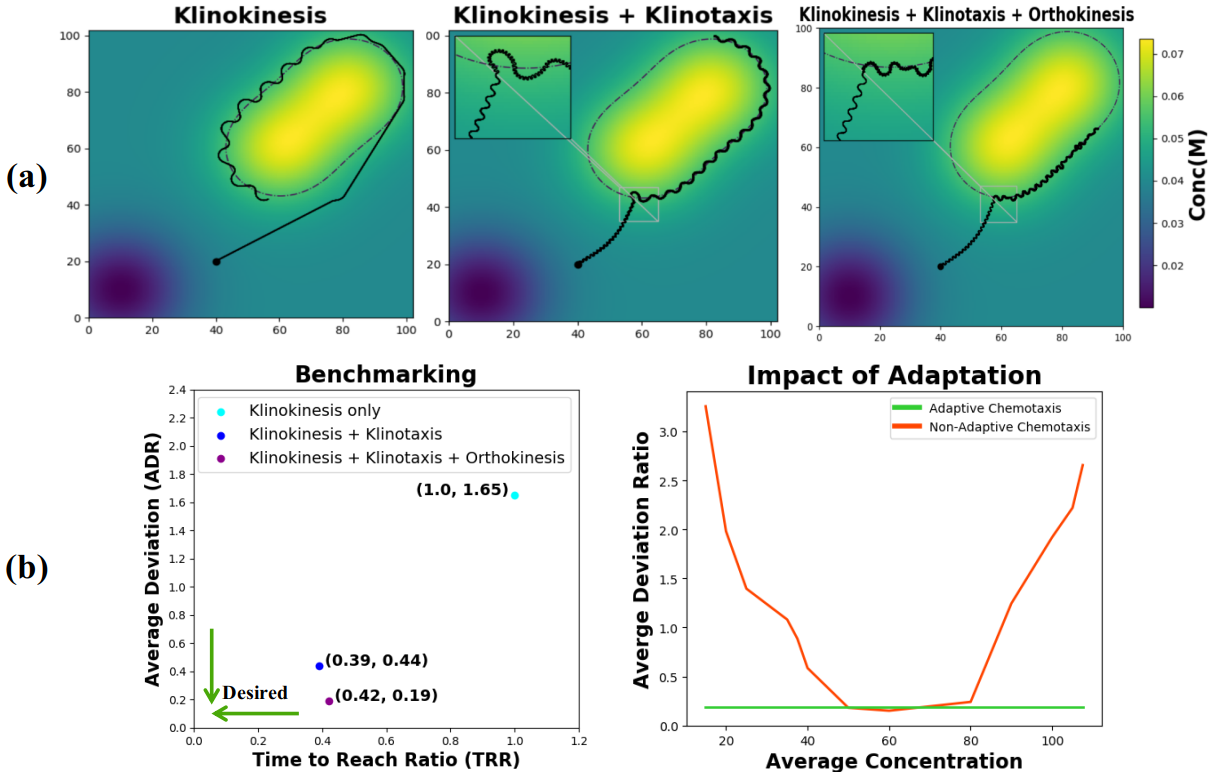}}
		\caption{(a) The dotted line in all three contour tracking plots is the setpoint contour corresponding to $C_T = 55mM$. (b) Left: Benchmarking with TRR and ADR. Right: Impact on ADR of dividing by $C_{avg}(t)$ while computing the gradient estimate.}
		\label{comp}
	\end{figure}

	The algorithms are visually compared in Fig.\ref{comp}(a). Note that the bot was started from the same starting point and initial angle in all three plots. Using only klinokinesis (left), the bot takes a long route to reach the setpoint and exhibits large overshoots around the setpoint contour. Adding klinotaxis (middle) allows the bot to reach the setpoint using a much faster route while also reducing setpoint deviation. Adding orthokinesis (right) further reduces the setpoint deviation. 
	
	We now define the Time to Reach Ratio (TRR) of an algorithm $A$ for a setpoint $C_T$ as the time taken to first reach $C_T$ using $A$ divided by the time to first reach $C_T$ using klinokinesis. Clearly, the TRR also depends on the particular concentration landscape, starting point and initial angle. Here we consider the aggregated TRR obtained by averaging the TRR over 10 landscapes, 10 starting points for each landscape and 10 initial angles for each tuple of landscape and starting point. Also note that the TRR for klinokinesis will trivially be $1.0$.
	The second metric is adopted from \cite{santurkar2015c,shukla2018design} to quantify the deviation from the setpoint once the bot has reached the contour. This metric is the average deviation ratio from setpoint (ADR), defined as $ADR = \frac{1}{T - T_0} \int_{T_0}^{T}\frac{\left | C(t) - C_T \right|}{C_{max} - C_{min}}dt$, where $T_0$ is the first time that the bot reaches $C_T$, $T$ is the total simulation time, $C_{max}$ and $C_{min}$ are respectively the maximum and minimum concentrations values on the landscape. Thus the ADR measures the time-averaged ratio of the absolute deviation to the landscape concentration range. We report the aggregated ADR by averaging over the same set of configurations as for the aggregrated TRR.
	
	The algorithms are benchmarked using these two metrics in the left panel of Fig. \ref{comp}(b). We find that there is a drastic reduction in the TRR, by a factor of 2.6, due to the inclusion of klinotaxis, implying that the bot reaches the setpoint faster.
	Remarkably this improvement is achieved despite the bot's effective speed being reduced by a factor of roughly 7.5 by moving along a sinusoidal path instead of a straight line. As expected, a second effect of this reduced effective velocity is that the inclusion of klinotaxis also reduces the ADR, by a significant factor of 3.8. 
	The TRR is slightly larger with the inclusion of orthokinesis as the bot slows down near the setpoint. However we observe a larger reduction in ADR, demonstrating that orthokinesis can adaptively trade-off speed for a significant reduction in setpoint overshoot. The TRR and ADR respectively reduced by a factor of 2.4 and 8.7 by including both klinotaxis and orthokinesis (w.r.t just klinokinesis). Also note that the standard deviation of the TRR for both the ``klinokinesis + klinotaxis'' and ``klinokinesis + klinotaxis + orthokinesis'' algorithms were found to be $0.05$. The standard deviation of the ADR for ``klinokinesis only'', ``klinokinesis + klinotaxis'' and ``klinokinesis + klinotaxis + orthokinesis'' algorithms were respectively found to be $0.34$, $0.08$ and $0.04$. 
	
	In the right panel of Fig.\ref{comp}(b), we quantitatively demonstrate the drastic improvement in robustness of chemotaxis due to the inclusion of $C_{avg}(t)$ in the denominator of (1). We plot the ADR for the ``klinokinesis + klinotaxis + orthokinesis'' algorithm as a function of average landscape concentration. Without adaptive scaling (red), the ADR is comparable to that with adaptive scaling (green) in a narrow range of average concentration, but degrades rapidly away from this optimal range. Similar plots were also obtained for the ``klinokinesis only'' and ``klinokinesis + klinotaxis'' algorithms. 
	This shows that previously proposed chemotaxis algorithms in the literature (that did not incorporate dynamic scaling) are not robust to concenctration rescaling, highlighting the importance of the novel dynamic scaling proposed in this paper. Finally, we refer the reader to \cite{santurkar2015c} for a demonstration of the SNN-based klinokinesis-only strategy achieving lower ADR compared to PID control while also being significantly more energy efficient, with these results holding transitively for the schemes proposed here.
	
	In conclusion, we have presented a scale-invariant, adaptive chemotaxis algorithm using spiking neurons that successfully combines klinotaxis, klinokinesis and orthokinesis. This allows us to perform robust, resource constrained and energy efficient contour tracking while achieving state-of-the-art performance.

	\bibliographystyle{splncs04}
	\bibliography{ref}
	
\end{document}